\documentclass[final,5p,times,twocolumn]{elsarticle}

\usepackage{graphicx}

\usepackage{amssymb}
\usepackage[utf8]{inputenc}
\usepackage{units}

\journal{NIM A}

\newcommand{\xmax}{X$_{\mathrm{max}}$ }

\begin{document}


\begin{frontmatter}

\title{Radio detection of extensive air showers}

\author[KITIK]{Tim Huege}

\address[KITIK]{Institut f\"ur Kernphysik, Karlsruher Institut f\"ur Technologie - Campus Nord, Postfach 3640, 76021 Karlsruhe, Germany}

\cortext[cor]{Email: tim.huege@kit.edu}

\begin{abstract}
Radio detection of extensive air showers initiated in the Earth's atmosphere has made tremendous progress in the last decade. Today, radio detection is routinely used in several cosmic-ray observatories. The physics of the radio emission in air showers is well-understood, and analysis techniques have been developed to determine the arrival direction, the energy and an estimate for the mass of the primary particle from the radio measurements. The achieved resolutions are competitive with those of more traditional techniques. In this article, I shortly review the most important achievements and discuss the potential for future applications.
\end{abstract}

\begin{keyword}

high-energy cosmic rays \sep radio emission \sep extensive air showers


\end{keyword}

\end{frontmatter}




\section{Introduction}

\begin{figure*}[!htb]
  \centering
  \includegraphics[width=0.63\textwidth]{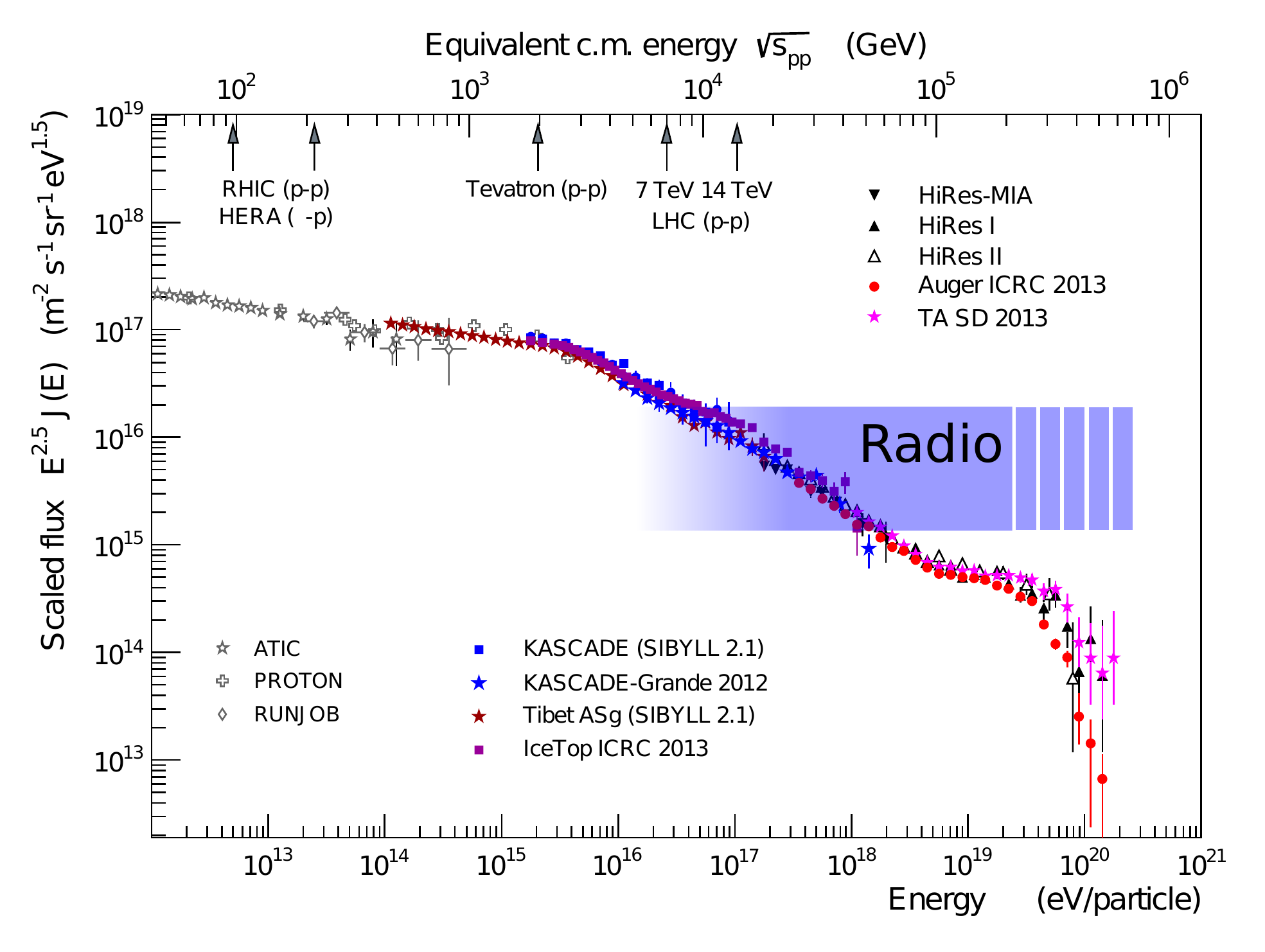}
  \caption{Cosmic-ray energy spectrum overlaid with the reach of the VHF radio detection technique.
  At low energies the radio signals are hidden in Galactic noise, at very high energies concepts
  have yet to be devised to cover extremely large detection areas. Diagram updated and adapted from \citep{Engel:2011zzb}, reprinted from \citep{HuegePLREP}.}
  \label{fig:crspectrum}
 \end{figure*}

While the understanding of cosmic rays has progressed very significantly in the last decades, many questions about their origin and the physics of their acceleration are still unanswered \citep{Bluemer2009293}. New detection techniques help to maximize the information gathered about each detected cosmic-ray particle. This is especially important at the highest energies where fluxes are extremely low and measurements through air showers yield only rather indirect information on the primary particles. In the last years, radio detection of air showers in the \emph{very high frequency} (VHF) band, typically around \unit[30--80]{MHz}, has been researched with great effort. Today, the technique has matured from a prototype stage to a well-established detection technique that benefits any cosmic-ray detector in which it is employed. The energy range accessible with the so-far developed approaches is illustrated in Fig.\ \ref{fig:crspectrum}

In the following, I give a concise overview of the most important achievements made with the technique to date. For a detailed discussion of the state of the field, I kindly refer the reader to a previously-published extensive review \citep{HuegePLREP}.

\section{Emission physics}

The most important breakthrough of the past few years has been the detailed understanding of the radio emission processes in extensive air showers. Three effects are important:
\begin{itemize}
\item Electrons and positrons in the extensive air shower are accelerated by the Lorentz force in the geomagnetic field. At the same time they are decelerated by interactions with air molecules. An equilibrium arises, and the net drift of the particles in directions perpendicular to the air-shower axis leads to transverse currents. As the shower first grows in particle number, then reaches a maximum and then dies out, these transverse currents undergo a time-variation. The time-variation of the currents leads to radio emission. This is the dominant effect responsible for approximately 90\% of the electric field amplitude, usually referred to as ``geomagnetic emission'' \citep{KahnLerche1966,ScholtenWernerRusydi2008}.
\item During the air-shower evolution, a negative charge excess builds up in the shower front. This arises mostly because ionization electrons from the ambient medium are swamped with the shower, while positive ions stay behind. Again, as the shower evolves, the net charge grows, reaches a maximum and then declines. The time-variation of the net charge excess leads to radio emission which contributes approximately 10\% of the electric field amplitude. This is the so-called ``Askaryan effect'' which is the dominant mechanism for radio emission from particle showers in dense media \citep{Askaryan1962a,Askaryan1965}.
\item At VHF frequencies the radio emission is generally coherent. This means that electric field amplitudes from individual particles add up constructively. The total electric field amplitude thus scales linearly with the number of particles in the air shower, which in turn scales approximately linearly with the energy of the primary cosmic ray. Consequently, the radiated power (and energy) scales quadratically with the cosmic-ray energy. Coherence is governed by different scales in the air shower: the thickness of the shower pancake, the lateral width of the shower, and the time-delays arising from the geometry of the shower disk propagating with the speed of light as seen from a specific observer location. The latter is strongly influenced by the refractive index of the air which is approximately 1.0003 at sea level and scales with the density gradient of the atmosphere. This leads to ``Cherenkov rings'' in the radio-emission footprints; observers on these rings see time-compressed radio signals for which coherence reaches up to GHz frequencies \citep{AlvarezMunizCarvalhoZas2012,DeVriesBergScholten2011}.
\end{itemize}

 \begin{figure*}[!htb]
  \centering
  \includegraphics[width=0.23\textwidth,clip=true,trim=0cm 0cm 0cm 12cm]{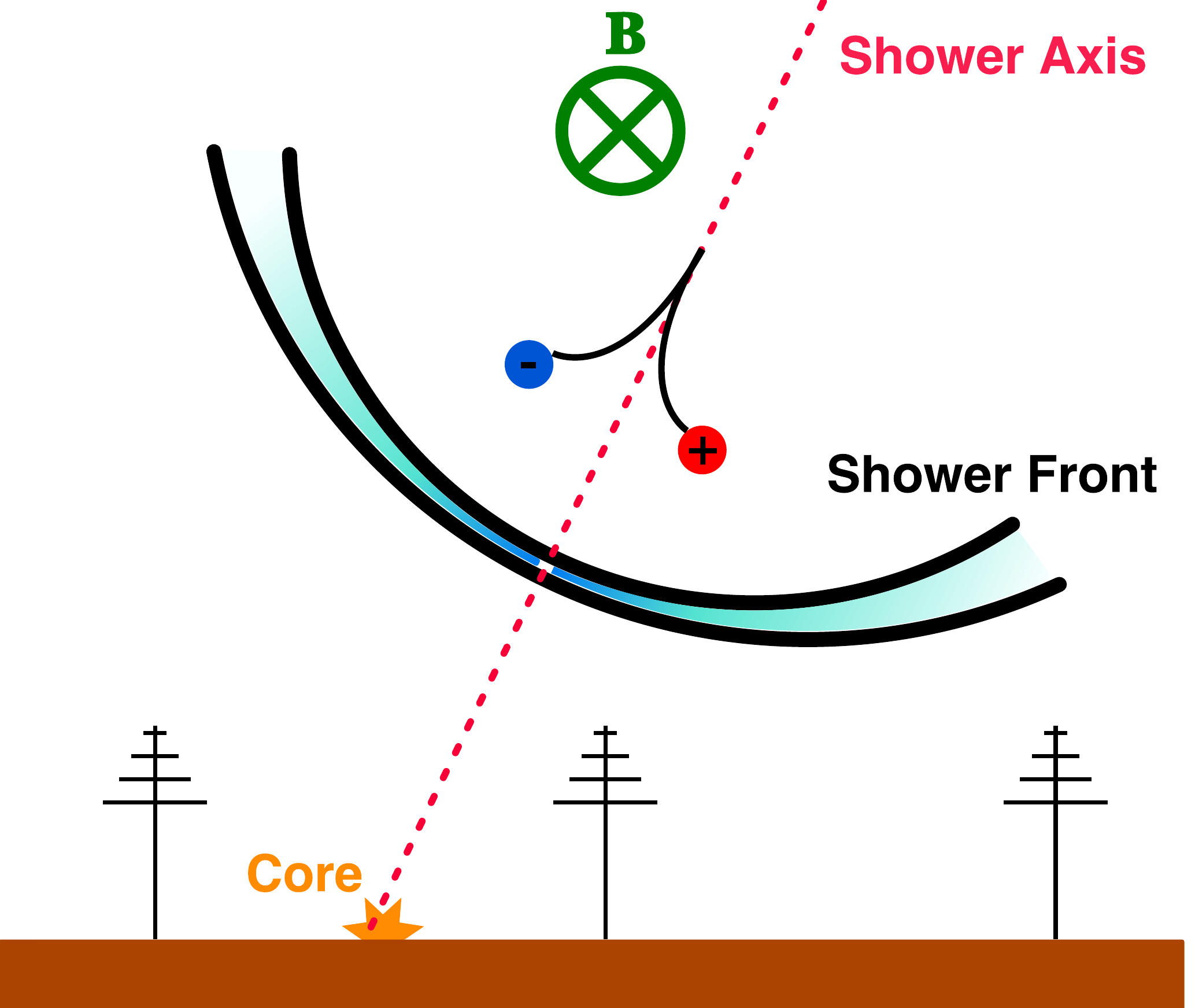}
  \includegraphics[width=0.21\textwidth]{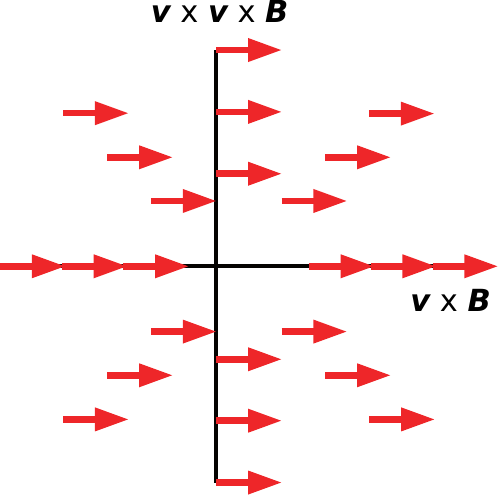}
  \hspace{0.06\textwidth}
  \includegraphics[width=0.23\textwidth,clip=true,trim=0cm 0cm 0cm 12cm]{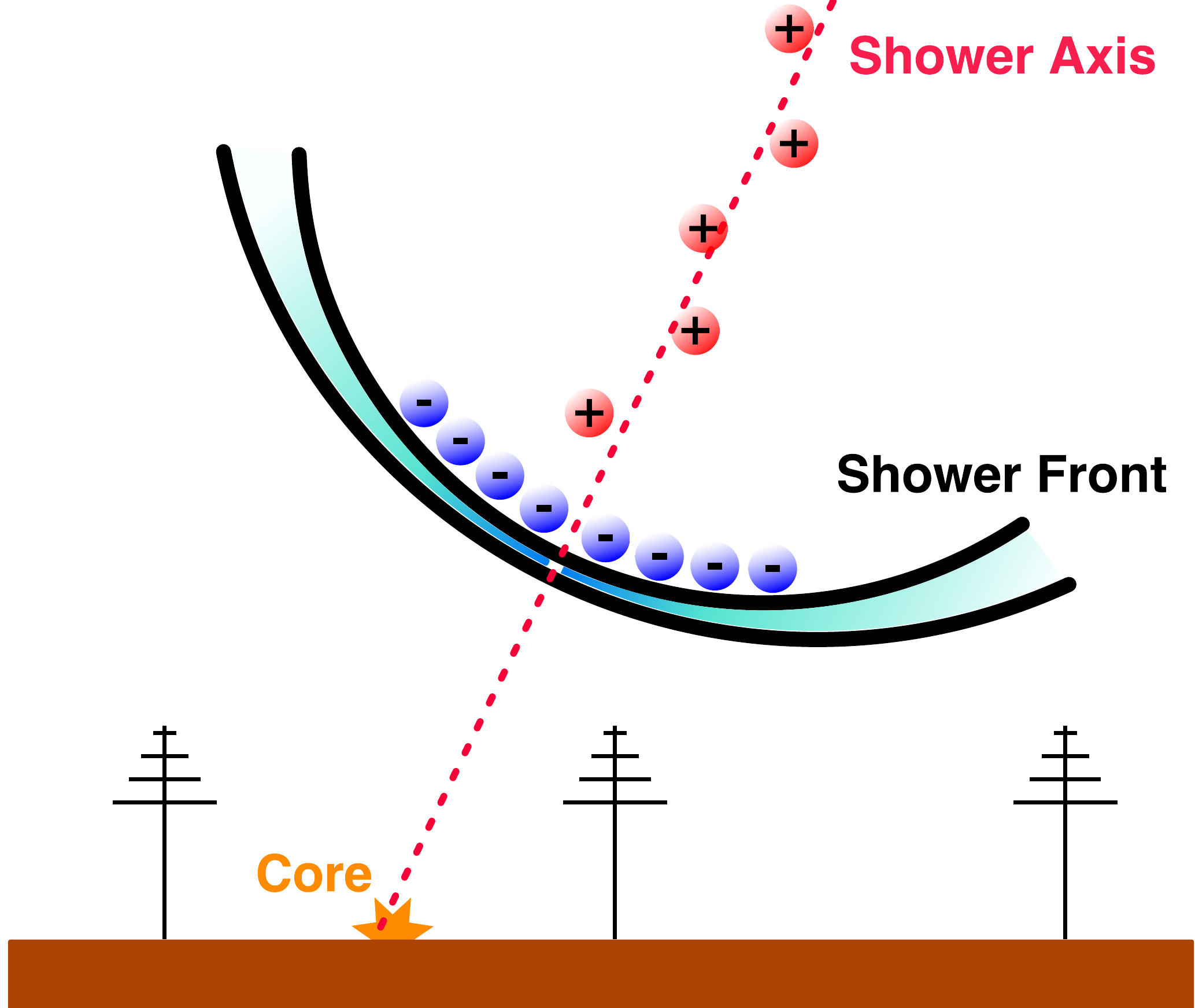}
  \includegraphics[width=0.21\textwidth]{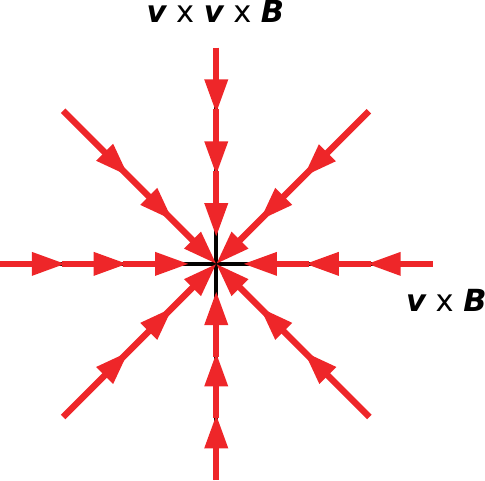}
  \caption{Left: The geomagnetic radiation mechanism. 
  The arrows indicate the direction of linear polarization in the plane 
  perpendicular to the air shower axis. The emission is linearly polarized along the direction 
  given by the Lorentz force, $\vec{v} \times \vec{B}$ (east-west for 
  vertical air showers). Right: 
  The charge excess (Askaryan) emission. The arrows 
  indicate the polarization which is linear with electric field vectors oriented
  radially with respect to the shower axis. Diagrams have been adapted from 
  \citep{SchoorlemmerThesis2012} and \citep{deVries2012S175} and reprinted from \citep{HuegePLREP}.}
  \label{fig:mechanisms}
 \end{figure*}

The geomagnetic and Askaryan mechanisms have different polarization characteristics, their superposition leads to constructive interference or destructive interference depending on the location of the observer with respect to the shower axis. The resulting radio-emission ``footprint'' is thus asymmetric. An illustration of the mechanisms and their polarization characteristics is shown in Fig.\ \ref{fig:mechanisms}.

\begin{figure}[t]
    \centering
    \includegraphics[width=0.45\textwidth]{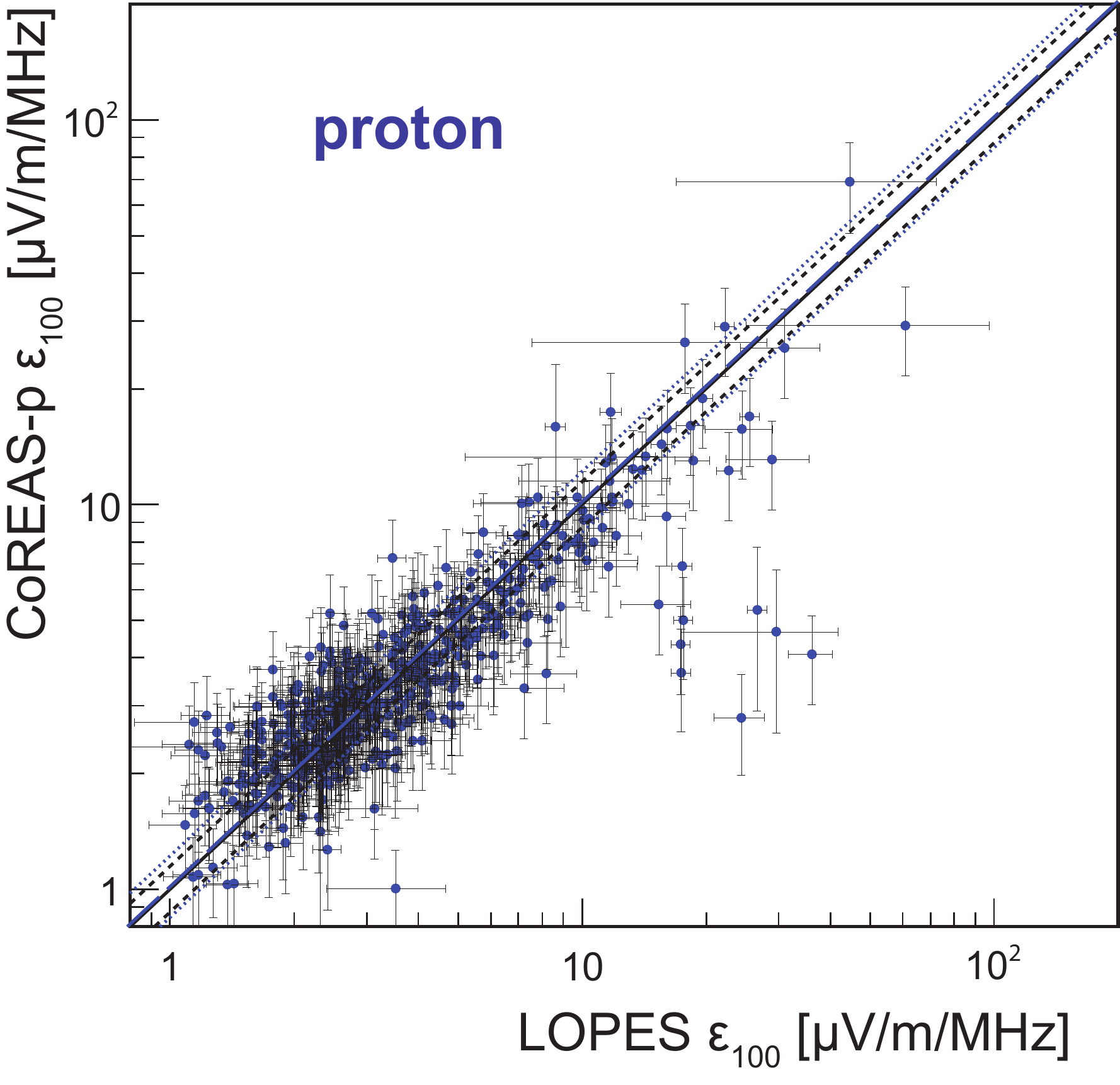}
    \caption{Comparison of the amplitude at a lateral 
    distance of 100~m as measured with LOPES and simulated with CoREAS for 
    air showers induced by protons. The few outlier events 
    in the lower-right parts of the diagrams 
    are not understood, but constitute less than 2\% of the data. Adapted from 
    \citep{HuegeLOPESIcrc2015}, reprinted from \citep{HuegePLREP}.}
    \label{fig:epscomparison}
\end{figure}

The interpretation discussed so far is based on macroscopic models which employ concepts such as currents and net charges in the air shower. Correct incorporation of the enormous complexity of the air shower and the resulting coherence effects in macroscopic models is, however, very difficult, and the required simplifications degrade the quality of the modelled radio signals. This is why microscopic simulations in which the radio emission from the particle shower is calculated by adding up the emission from each individual electron and positron \citep{AlvarezMunizCarvalhoZas2012,HuegeARENA2012a} are most widely used in the community. These are based on first principle calculations, applying discretized formalisms of classical electrodynamics \citep{ZasHalzenStanev1992,JamesFalckeHuege2012} to the individual moving particles in the air shower. Consequently, they predict the radio signal on an absolute scale without any free parameters in the simulation. All measurements to date have been described by such simulations within errors. An example for a comparison of amplitudes measured with LOPES \citep{FalckeNature2005} and simulated with CoREAS \citep{HuegeARENA2012a} at a distance of \unit[100]{m} from the shower axis is given in Fig.\ \ref{fig:epscomparison}. Similar results have been reported by other experiments.

\section{Reconstruction of cosmic-ray parameters}

Any detection technique for extensive air showers is only a means to the end of determining the parameters of the primary cosmic-ray particle: the arrival direction, the energy, and a measure for the mass. And in fact, all of these parameters have been demonstrated to be reconstructable from radio measurements with resolutions competitive with those of more traditional detection techniques.

The arrival direction of the air shower can be determined from the arrival-time distribution of the radio pulses in individual antennas. It is important to note that the wavefront of the radio signal is not a plane wave but has a complex structure: It is hyperbolical \citep{LOPESWavefront2014,LOFARWavefront2014}, i.e., spherical close to the shower axis and conical further away from the shower axis. The arrival direction has been demonstrated to be reconstructable within 0.5$^{\circ}$ and possibly as well as 0.1$^{\circ}$ with radio techniques.

\begin{figure}[!htb]
\centering
\includegraphics[width=0.45\textwidth]{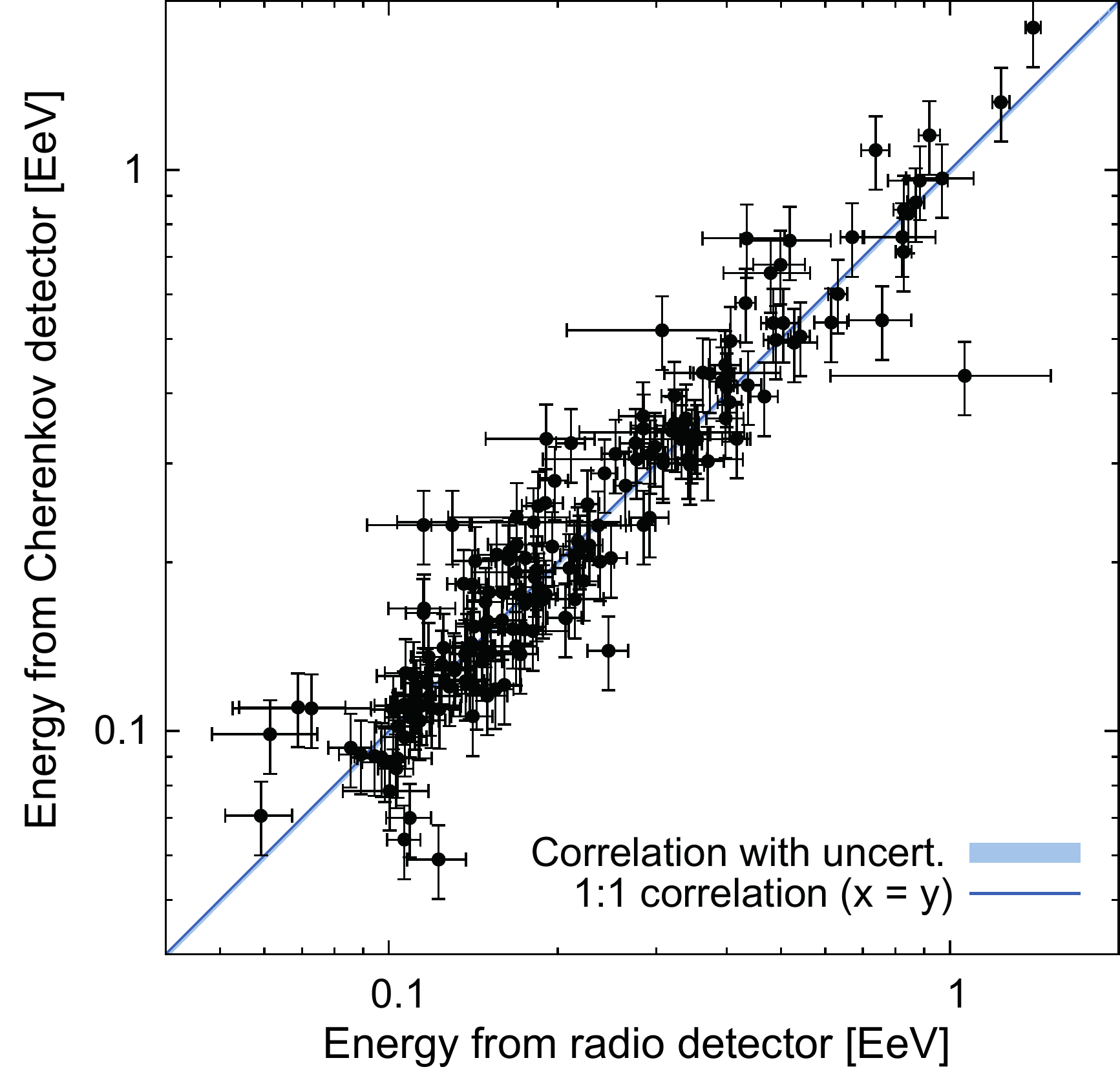}
\caption{Cosmic-ray energy determined with the Tunka-Rex radio antennas in comparison with the energy reconstructed with the 
Tunka-133 optical Cherenkov detectors. Adapted from 
\citep{TunkaRexCrossCalibration}, reprinted from \citep{HuegePLREP}.\label{fig:tunkarexenergy}}
\end{figure}

\begin{figure}[!htb]
\centering
\includegraphics[width=0.45\textwidth]{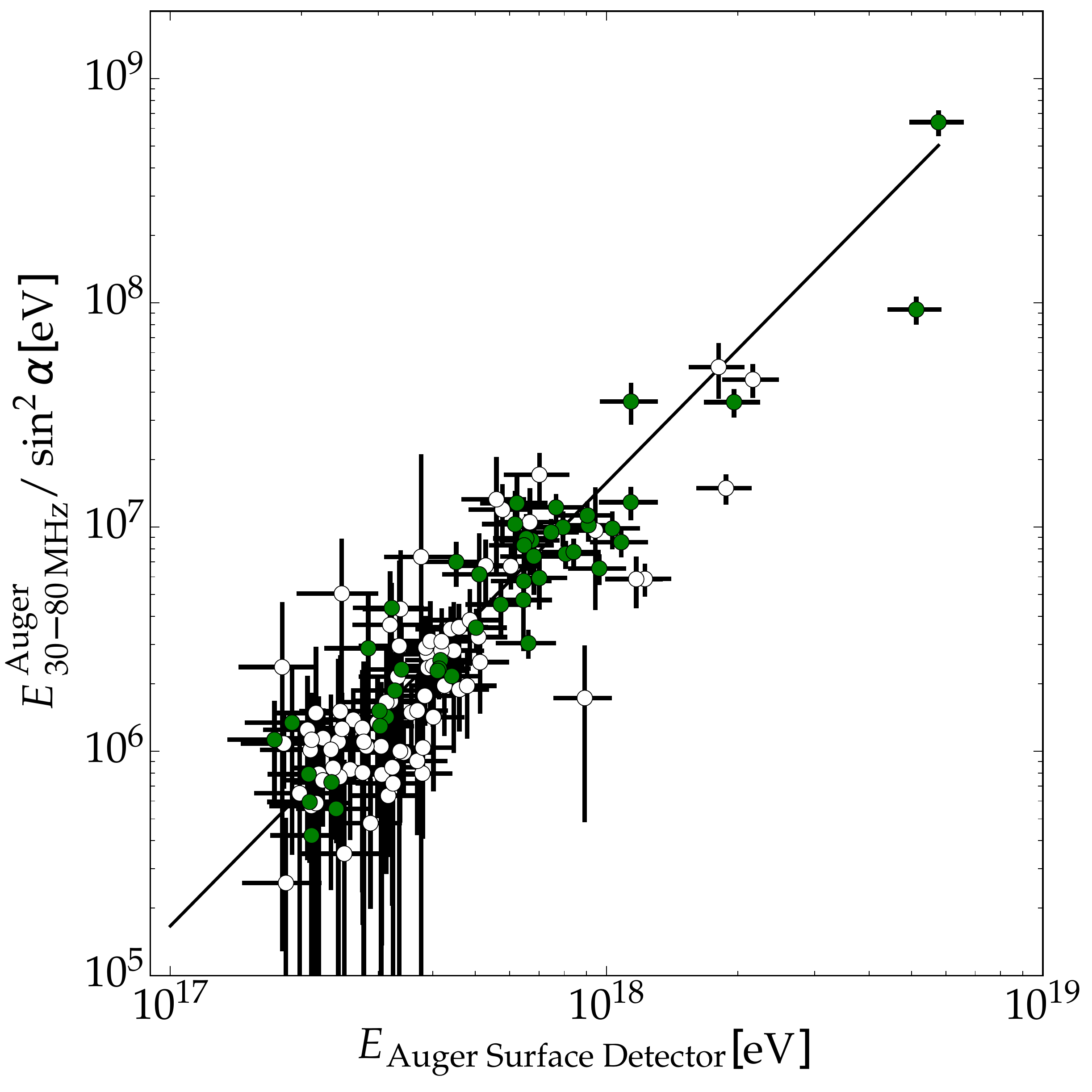}
\caption{Correlation between the radiation energy (normalized for incidence perpendicular to the geomagnetic field) and the cosmic-ray energy
measured with the Auger surface detector. Open circles represent air showers with radio signals 
measured in three or four AERA detectors, filled circles 
correspond to showers with five or more measured radio signals. Adapted 
from \citep{AERAEnergyPRD}, reprinted from \citep{HuegePLREP}.\label{fig:crosscalibration}}
\end{figure}

Due to the coherent nature of the radio emission, the amplitude of the radio signal scales approximately linearly with the energy of the primary particle. The depth of the shower maximum (related to the mass of the primary particle) influences the steepness of the lateral distribution of the radio signal, so a measurement at a given position will exhibit intrinsic fluctuations of the measured amplitude. Two concepts have been devised to minimize these intrinsic fluctuations for a precise measurement of the energy of the cosmic ray. First, a characteristic lateral distance exists \citep{HuegeUlrichEngel2008} at which these fluctuations are minimized. Experiments such as LOPES and Tunka-Rex \citep{TunkaRexInstrument} have thus used amplitude measurements at this characteristic lateral distance as an energy estimator, and have reached resolutions as good as 15\% using this approach, cf.\ Fig.\ \ref{fig:tunkarexenergy}. Second, instead of measuring the amplitude at a specific lateral distance, an area integral can be performed over the complete radio-emission footprint. This approach has been pioneered by the \emph{Auger Engineering Radio Array} (AERA) \citep{SchulzIcrc2015}. The energy fluence (in units of eV/m$^2$) as measured at individual antenna locations is fitted with a model of the two-dimensional emission footprint to then integrate over area and determine the total ``radiation energy'' (in units of eV) deposited in the form of radio signals on the ground. This radiation energy scales quadratically with the energy of the primary cosmic ray, the achieved energy resolution amounts to 17\%, cf.\ Fig.\ \ref{fig:crosscalibration}. The radiation energy has the benefit of being a well-defined physical quantity that is independent of the observation altitude and zenith angle of the air shower \citep{AERAEnergyPRL}. It is thus conceptually very attractive as a means to cross-calibrate the absolute energy scales of experiments against each other or against first-principle calculations. An interesting aside is that only a minute fraction of the energy of the primary particle is radiated in the form of radio signals in the VHF band: For a \unit[10$^{18}$]{eV} primary particle, the radiation energy amounts to approximately \unit[15.8]{MeV} \citep{AERAEnergyPRL}. Nevertheless, photon statistics need not be considered as the energy of a \unit[55]{MHz} photon is of order \unit[$10^{-7}$]{eV}.

\begin{figure}[!htb]
\centering
\includegraphics[width=0.48\textwidth]{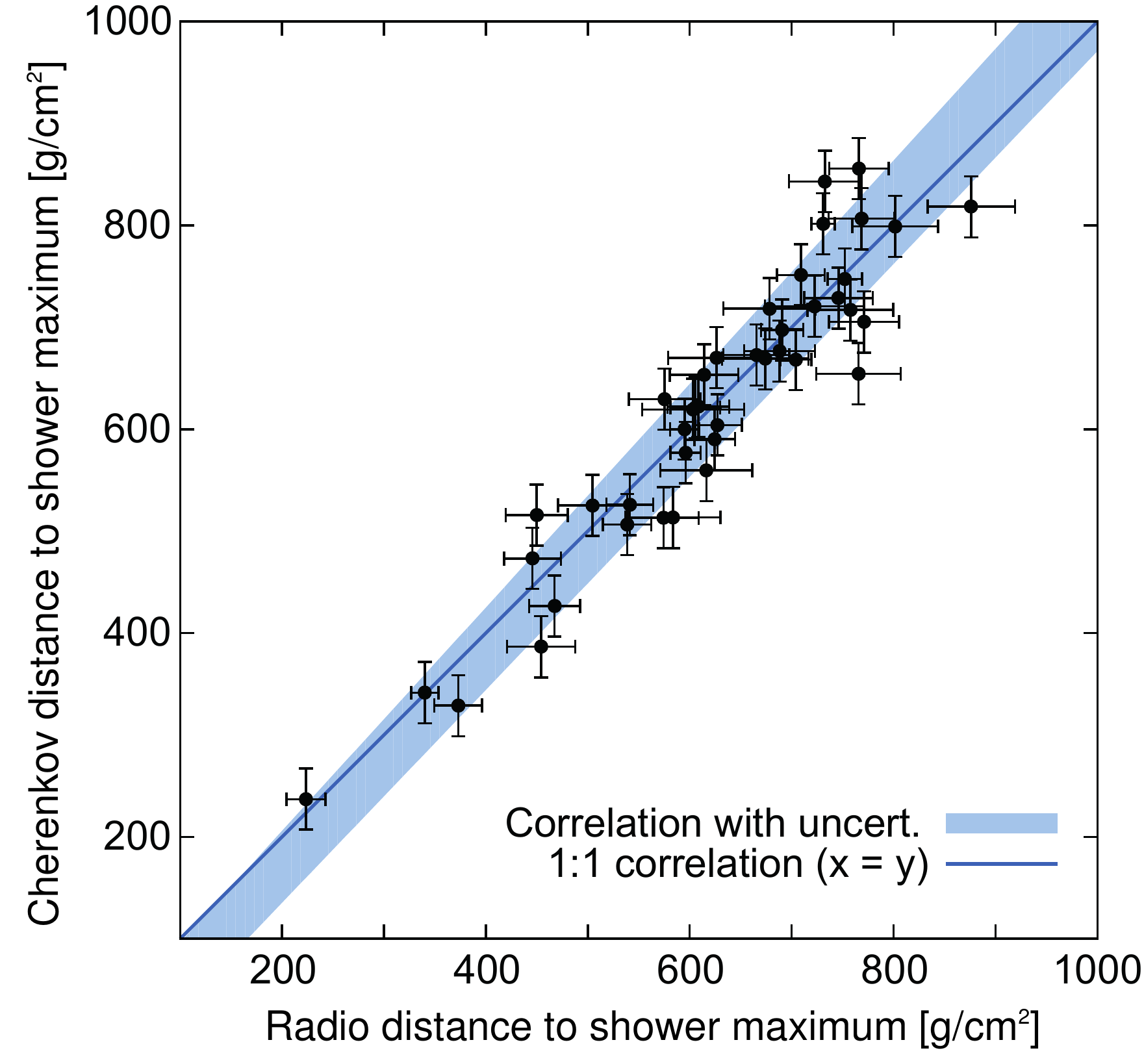}
\caption{Atmospheric depth between shower maximum and observer altitude as determined with the Tunka-Rex radio measurement and 
the Tunka-133 Cherenkov detectors. Adapted from \citep{TunkaRexCrossCalibration}, reprinted from \citep{HuegePLREP}.\label{fig:tunkarexxmax}}
\end{figure}

\begin{figure*}[htb]
  \includegraphics[width=0.32\textwidth]{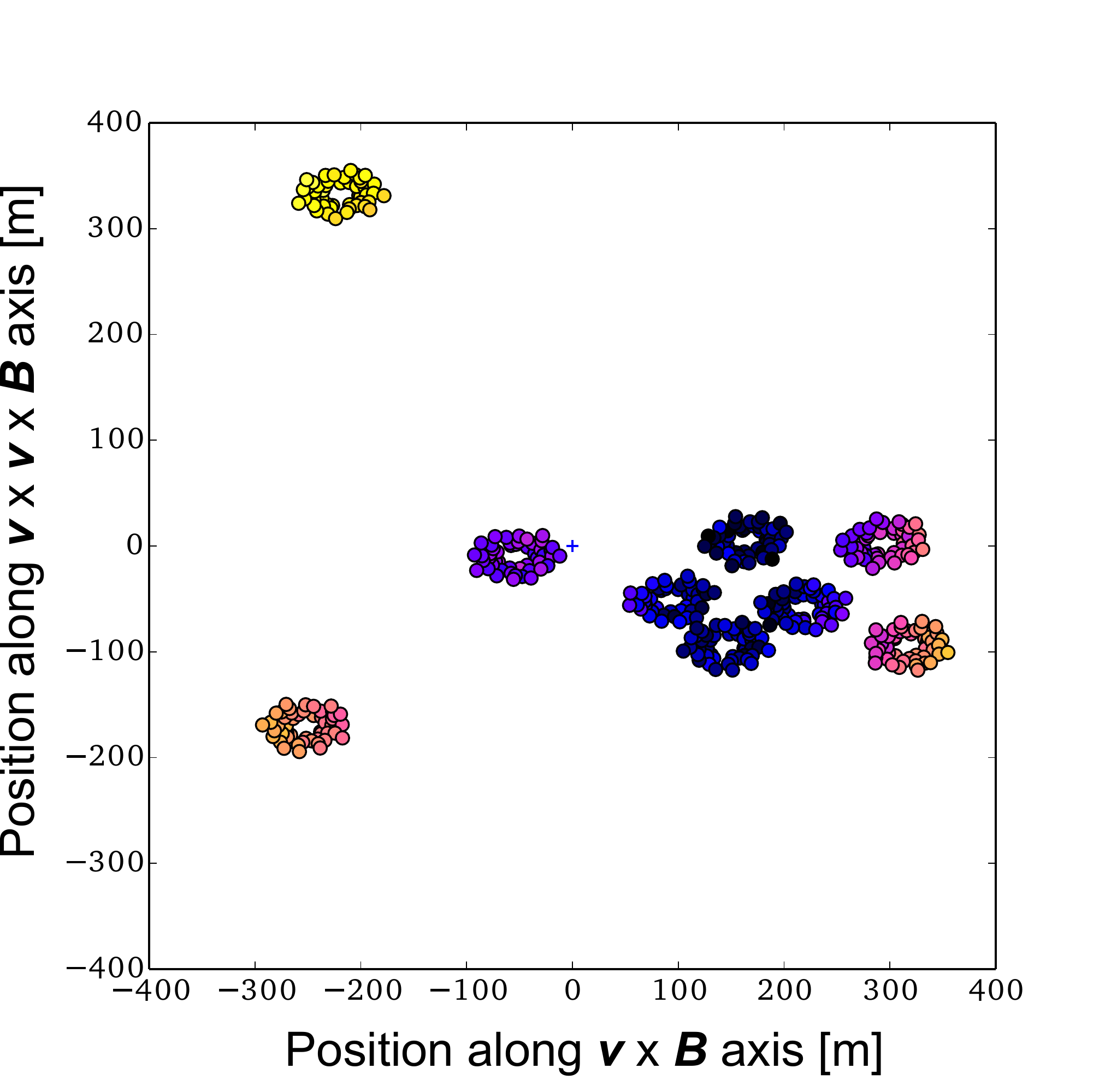}
  \includegraphics[clip=true,trim=50 0 45 0,width=0.35\textwidth]{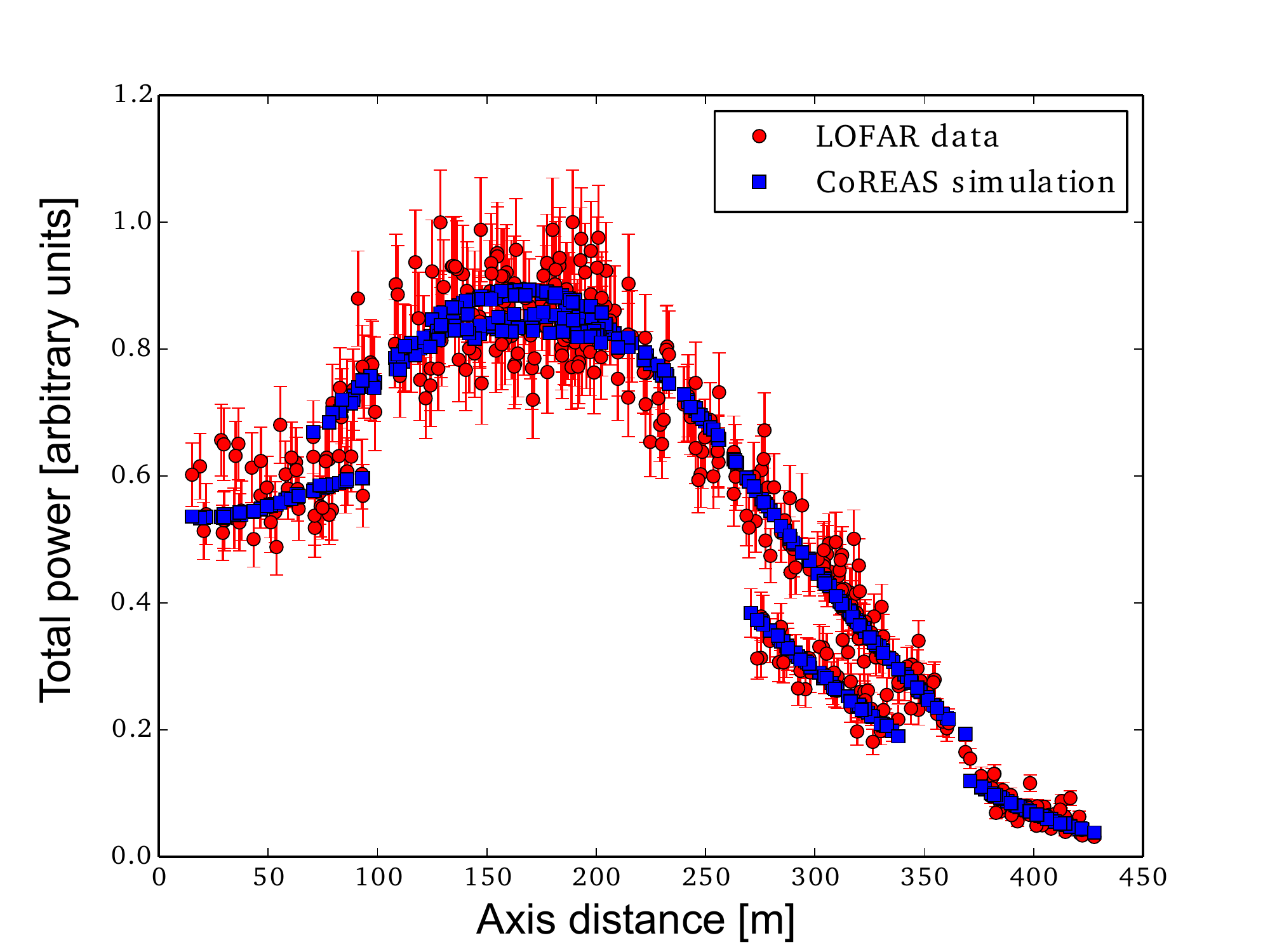}
  \includegraphics[width=0.31\textwidth]{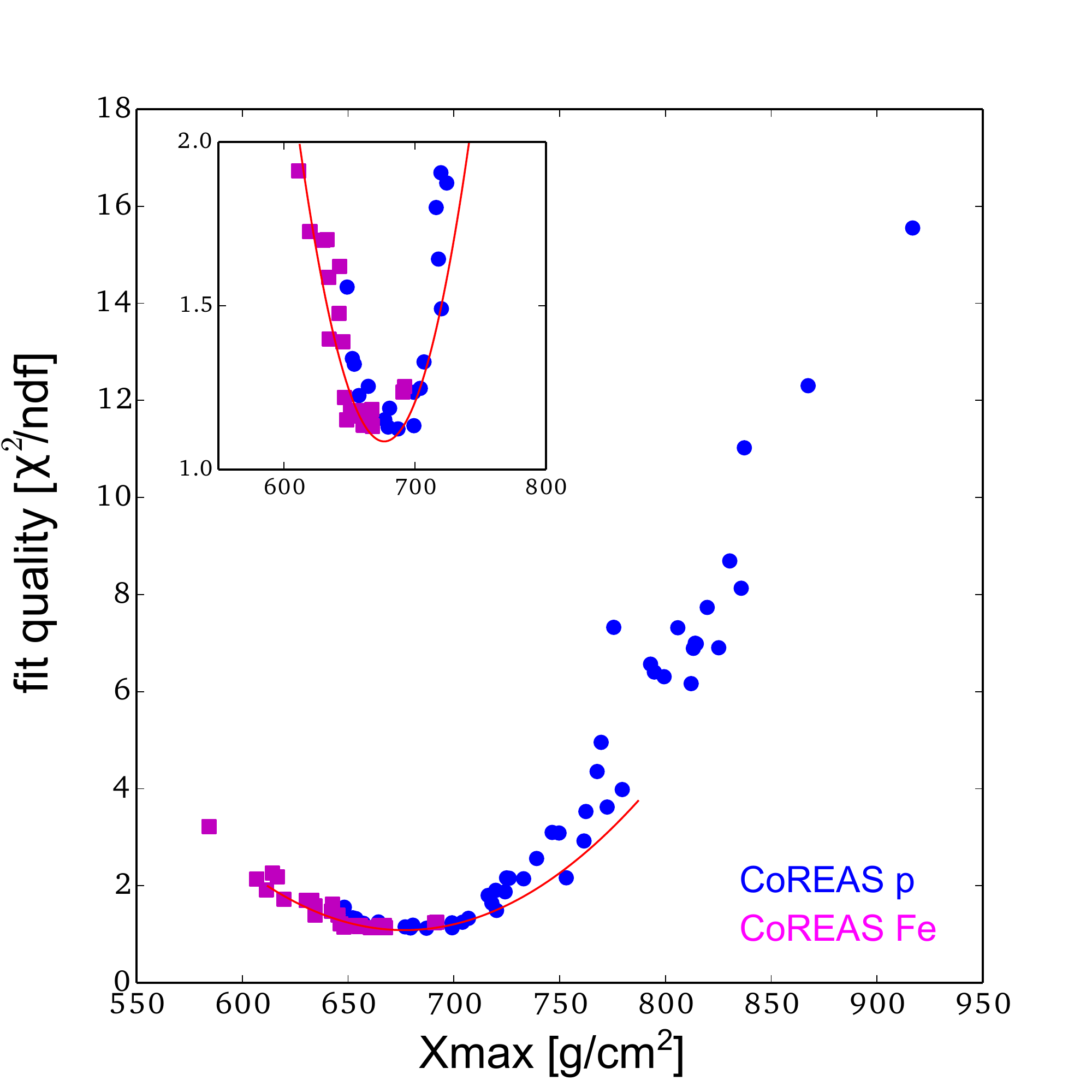}
  \caption{Left: Energy fluence measured at individual LOFAR 
  antennas (colored circles) in comparison with the  
  signal distribution predicted by the best-fitting of a set of 
  CoREAS simulations (background-color) for a particular air shower 
  event. Middle: One-dimensional projection of the 
  two-dimensional signal distribution. Right: Quality of the 
  agreement between the energy fluence distribution measured with LOFAR 
  and the distribution predicted by different CoREAS simulations of the air 
  shower event. A clear correlation between the value of \xmax and the quality of the fit is 
  obvious. All diagrams adapted from \citep{LOFARXmaxMethod2014}, reprinted from \citep{HuegePLREP}.}\label{fig:xmaxlofar}
\end{figure*}

The most difficult challenge is the reconstruction of the depth of shower maximum (in g/cm$^2$, usually called $X_{\mathrm{max}}$), which is the primary estimator for particle mass used in air-shower measurements. Fluorescence and Cherenkov light detectors are able to measure \xmax with a resolution of \unit[20--25]{g/cm$^2$}. Radio signals from air showers carry information on the distance of the source of the emission and thus on the depth of the shower maximum, encoded in the steepness of the lateral amplitude distribution as well as in the wavefront structure and the radio pulse shapes. So far, most analyses only exploit the lateral signal distribution. The Tunka-Rex experiment has demonstrated a clear correlation between \xmax measured with the Tunka Cherenkov light detectors and the Tunka-Rex radio antennas, cf.\ Fig.\ \ref{fig:tunkarexxmax}. The estimated \xmax resolution of the radio reconstruction is of order \unit[40]{g/cm$^2$} and thus not yet competitive with established techniques. However, there is still room for improvement, e.g., by exploiting additional signal information. Using the dense antenna array of LOFAR \citep{LOFARCosmicRays} and a top-down approach where individual events are compared with dozens of simulations to determine the true \xmax value from the best-fitting simulation (see Fig.\ \ref{fig:xmaxlofar}), \xmax resolutions better than \unit[20]{g/cm$^2$} are achievable also with radio measurements. With such a good measurement precision, even small data sets can be used to begin constrain the mass composition of cosmic rays \citep{LOFARNatureXmax}.

\section{Future applications}

There are numerous applications in which radio detectors can benefit existing or newly built cosmic-ray detectors. In particular:
\begin{itemize}
\item Any particle detector will profit from additional radio antennas which allow measurements of the pure electromagnetic cascade with a very good energy resolution and reasonable \xmax resolution. This is attractive in particular if much of the infrastructure (cabling, power, communications) can be shared between the detectors --- the radio antenna and readout electronics themselves are fairly inexpensive with prices below 1,000~USD per antenna certainly achievable.
\item Radio detectors can be used for a precise calibration of the absolute energy scale of a cosmic-ray detector. This is because the radio signal is not influenced by atmospheric conditions (no scattering, no absorption) and gives direct access to the calorimetric energy in the electromagnetic cascade of the air shower \citep{GlaserRadEnergyStudy}. Cross-calibration between different experiments, e.g., via the radiation energy, or even calibration of experiments using first-principle calculations are thus very attractive.
\item Horizontal air showers illuminate areas of several km$^2$ \citep{KambeitzARENA2016} and can thus be detected with antenna arrays using grid spacings of a kilometer or sparser, meaning that measurements of inclined air showers can be used to measure the electromagnetic component of ultra-high-energy cosmic rays.
\item Very dense antenna arrays might allow cosmic-ray measurements with unprecedented reconstruction quality for individual air-shower events. The upcoming Square Kilometre Array is expected to reach \xmax resolutions below \unit[10]{g/cm$^2$} \citep{AnneSKAARENA2016} and could investigate the mass composition in the region of transition from Galactic to extragalactic cosmic rays with unprecedented mass resolution.
\end{itemize}

\section{Conclusions}

Over the past decade, radio detection of extensive air showers has matured from a prototype phase with small installations and an unclear picture of the radio emission mechanisms to full-fledged detector arrays and a detailed understanding of the radio emission physics. Radio detection has by now become a routine part of several cosmic-ray detection efforts and contributes valuable information for the analysis of cosmic-ray data. Particular potential lies in a precise calibration of the absolute energy scale of cosmic-ray detectors, in the large-scale detection of horizontal air showers and in precision measurements with very dense arrays.


\end{document}